\documentstyle[12pt]{article}
\newcommand{\sh}{\sinh}
\newcommand{\bea}{\begin{eqnarray}}
\newcommand{\eea}{\end{eqnarray}}
\newcommand{\q}{\theta}
\begin{document}

\immediate\write16{<<WARNING: LINEDRAW macros work with emTeX-dvivers
                    and other drivers supporting emTeX \special's
                    (dviscr, dvihplj, dvidot, dvips, dviwin, etc.) >>}

\newdimen\Lengthunit       \Lengthunit  = 1.5cm
\newcount\Nhalfperiods     \Nhalfperiods= 9
\newcount\magnitude        \magnitude = 1000

\catcode`\*=11
\newdimen\L*   \newdimen\d*   \newdimen\d**
\newdimen\dm*  \newdimen\dd*  \newdimen\dt*
\newdimen\a*   \newdimen\b*   \newdimen\c*
\newdimen\a**  \newdimen\b**
\newdimen\xL*  \newdimen\yL*
\newdimen\rx*  \newdimen\ry*
\newdimen\tmp* \newdimen\linwid*

\newcount\k*   \newcount\l*   \newcount\m*
\newcount\k**  \newcount\l**  \newcount\m**
\newcount\n*   \newcount\dn*  \newcount\r*
\newcount\N*   \newcount\*one \newcount\*two  \*one=1 \*two=2
\newcount\*ths \*ths=1000
\newcount\angle*  \newcount\q*  \newcount\q**
\newcount\angle** \angle**=0
\newcount\sc*     \sc*=0

\newtoks\cos*  \cos*={1}
\newtoks\sin*  \sin*={0}

\catcode`\[=13

\def\rotate(#1){\advance\angle**#1\angle*=\angle**
\q**=\angle*\ifnum\q**<0\q**=-\q**\fi
\ifnum\q**>360\q*=\angle*\divide\q*360\multiply\q*360\advance\angle*-\q*\fi
\ifnum\angle*<0\advance\angle*360\fi\q**=\angle*\divide\q**90\q**=\q**
\def\sgcos*{+}\def\sgsin*{+}\relax
\ifcase\q**\or
 \def\sgcos*{-}\def\sgsin*{+}\or
 \def\sgcos*{-}\def\sgsin*{-}\or
 \def\sgcos*{+}\def\sgsin*{-}\else\fi
\q*=\q**
\multiply\q*90\advance\angle*-\q*
\ifnum\angle*>45\sc*=1\angle*=-\angle*\advance\angle*90\else\sc*=0\fi
\def[##1,##2]{\ifnum\sc*=0\relax
\edef\cs*{\sgcos*.##1}\edef\sn*{\sgsin*.##2}\ifcase\q**\or
 \edef\cs*{\sgcos*.##2}\edef\sn*{\sgsin*.##1}\or
 \edef\cs*{\sgcos*.##1}\edef\sn*{\sgsin*.##2}\or
 \edef\cs*{\sgcos*.##2}\edef\sn*{\sgsin*.##1}\else\fi\else
\edef\cs*{\sgcos*.##2}\edef\sn*{\sgsin*.##1}\ifcase\q**\or
 \edef\cs*{\sgcos*.##1}\edef\sn*{\sgsin*.##2}\or
 \edef\cs*{\sgcos*.##2}\edef\sn*{\sgsin*.##1}\or
 \edef\cs*{\sgcos*.##1}\edef\sn*{\sgsin*.##2}\else\fi\fi
\cos*={\cs*}\sin*={\sn*}\global\edef\gcos*{\cs*}\global\edef\gsin*{\sn*}}\relax
\ifcase\angle*[9999,0]\or
[999,017]\or[999,034]\or[998,052]\or[997,069]\or[996,087]\or
[994,104]\or[992,121]\or[990,139]\or[987,156]\or[984,173]\or
[981,190]\or[978,207]\or[974,224]\or[970,241]\or[965,258]\or
[961,275]\or[956,292]\or[951,309]\or[945,325]\or[939,342]\or
[933,358]\or[927,374]\or[920,390]\or[913,406]\or[906,422]\or
[898,438]\or[891,453]\or[882,469]\or[874,484]\or[866,499]\or
[857,515]\or[848,529]\or[838,544]\or[829,559]\or[819,573]\or
[809,587]\or[798,601]\or[788,615]\or[777,629]\or[766,642]\or
[754,656]\or[743,669]\or[731,681]\or[719,694]\or[707,707]\or
\else[9999,0]\fi}

\catcode`\[=12

\def\GRAPH(hsize=#1)#2{\hbox to #1\Lengthunit{#2\hss}}

\def\Linewidth#1{\global\linwid*=#1\relax
\global\divide\linwid*10\global\multiply\linwid*\mag
\global\divide\linwid*100\special{em:linewidth \the\linwid*}}

\Linewidth{.4pt}
\def\sm*{\special{em:moveto}}
\def\sl*{\special{em:lineto}}
\let\moveto=\sm*
\let\lineto=\sl*
\newbox\spm*   \newbox\spl*
\setbox\spm*\hbox{\sm*}
\setbox\spl*\hbox{\sl*}

\def\mov#1(#2,#3)#4{\rlap{\L*=#1\Lengthunit
\xL*=#2\L* \yL*=#3\L*
\xL*=\xscale\xL* \yL*=\yscale\yL*
\rx* \the\cos*\xL* \tmp* \the\sin*\yL* \advance\rx*-\tmp*
\ry* \the\cos*\yL* \tmp* \the\sin*\xL* \advance\ry*\tmp*
\kern\rx*\raise\ry*\hbox{#4}}}

\def\rmov*(#1,#2)#3{\rlap{\xL*=#1\yL*=#2\relax
\rx* \the\cos*\xL* \tmp* \the\sin*\yL* \advance\rx*-\tmp*
\ry* \the\cos*\yL* \tmp* \the\sin*\xL* \advance\ry*\tmp*
\kern\rx*\raise\ry*\hbox{#3}}}

\def\lin#1(#2,#3){\rlap{\sm*\mov#1(#2,#3){\sl*}}}

\def\arr*(#1,#2,#3){\rmov*(#1\dd*,#1\dt*){\sm*
\rmov*(#2\dd*,#2\dt*){\rmov*(#3\dt*,-#3\dd*){\sl*}}\sm*
\rmov*(#2\dd*,#2\dt*){\rmov*(-#3\dt*,#3\dd*){\sl*}}}}

\def\arrow#1(#2,#3){\rlap{\lin#1(#2,#3)\mov#1(#2,#3){\relax
\d**=-.012\Lengthunit\dd*=#2\d**\dt*=#3\d**
\arr*(1,10,4)\arr*(3,8,4)\arr*(4.8,4.2,3)}}}

\def\arrlin#1(#2,#3){\rlap{\L*=#1\Lengthunit\L*=.5\L*
\lin#1(#2,#3)\rmov*(#2\L*,#3\L*){\arrow.1(#2,#3)}}}

\def\dasharrow#1(#2,#3){\rlap{{\Lengthunit=0.9\Lengthunit
\dashlin#1(#2,#3)\mov#1(#2,#3){\sm*}}\mov#1(#2,#3){\sl*
\d**=-.012\Lengthunit\dd*=#2\d**\dt*=#3\d**
\arr*(1,10,4)\arr*(3,8,4)\arr*(4.8,4.2,3)}}}

\def\clap#1{\hbox to 0pt{\hss #1\hss}}

\def\ind(#1,#2)#3{\rlap{\L*=.1\Lengthunit
\xL*=#1\L* \yL*=#2\L*
\rx* \the\cos*\xL* \tmp* \the\sin*\yL* \advance\rx*-\tmp*
\ry* \the\cos*\yL* \tmp* \the\sin*\xL* \advance\ry*\tmp*
\kern\rx*\raise\ry*\hbox{\lower2pt\clap{$#3$}}}}

\def\sh*(#1,#2)#3{\rlap{\dm*=\the\n*\d**
\xL*=\xscale\dm* \yL*=\yscale\dm* \xL*=#1\xL* \yL*=#2\yL*
\rx* \the\cos*\xL* \tmp* \the\sin*\yL* \advance\rx*-\tmp*
\ry* \the\cos*\yL* \tmp* \the\sin*\xL* \advance\ry*\tmp*
\kern\rx*\raise\ry*\hbox{#3}}}

\def\calcnum*#1(#2,#3){\a*=1000sp\b*=1000sp\a*=#2\a*\b*=#3\b*
\ifdim\a*<0pt\a*-\a*\fi\ifdim\b*<0pt\b*-\b*\fi
\ifdim\a*>\b*\c*=.96\a*\advance\c*.4\b*
\else\c*=.96\b*\advance\c*.4\a*\fi
\k*\a*\multiply\k*\k*\l*\b*\multiply\l*\l*
\m*\k*\advance\m*\l*\n*\c*\r*\n*\multiply\n*\n*
\dn*\m*\advance\dn*-\n*\divide\dn*2\divide\dn*\r*
\advance\r*\dn*
\c*=\the\Nhalfperiods5sp\c*=#1\c*\ifdim\c*<0pt\c*-\c*\fi
\multiply\c*\r*\N*\c*\divide\N*10000}

\def\dashlin#1(#2,#3){\rlap{\calcnum*#1(#2,#3)\relax
\d**=#1\Lengthunit\ifdim\d**<0pt\d**-\d**\fi
\divide\N*2\multiply\N*2\advance\N*\*one
\divide\d**\N*\sm*\n*\*one\sh*(#2,#3){\sl*}\loop
\advance\n*\*one\sh*(#2,#3){\sm*}\advance\n*\*one
\sh*(#2,#3){\sl*}\ifnum\n*<\N*\repeat}}

\def\dashdotlin#1(#2,#3){\rlap{\calcnum*#1(#2,#3)\relax
\d**=#1\Lengthunit\ifdim\d**<0pt\d**-\d**\fi
\divide\N*2\multiply\N*2\advance\N*1\multiply\N*2\relax
\divide\d**\N*\sm*\n*\*two\sh*(#2,#3){\sl*}\loop
\advance\n*\*one\sh*(#2,#3){\kern-1.48pt\lower.5pt\hbox{\rm.}}\relax
\advance\n*\*one\sh*(#2,#3){\sm*}\advance\n*\*two
\sh*(#2,#3){\sl*}\ifnum\n*<\N*\repeat}}

\def\shl*(#1,#2)#3{\kern#1#3\lower#2#3\hbox{\unhcopy\spl*}}

\def\trianglin#1(#2,#3){\rlap{\toks0={#2}\toks1={#3}\calcnum*#1(#2,#3)\relax
\dd*=.57\Lengthunit\dd*=#1\dd*\divide\dd*\N*
\divide\dd*\*ths \multiply\dd*\magnitude
\d**=#1\Lengthunit\ifdim\d**<0pt\d**-\d**\fi
\multiply\N*2\divide\d**\N*\sm*\n*\*one\loop
\shl**{\dd*}\dd*-\dd*\advance\n*2\relax
\ifnum\n*<\N*\repeat\n*\N*\shl**{0pt}}}

\def\wavelin#1(#2,#3){\rlap{\toks0={#2}\toks1={#3}\calcnum*#1(#2,#3)\relax
\dd*=.23\Lengthunit\dd*=#1\dd*\divide\dd*\N*
\divide\dd*\*ths \multiply\dd*\magnitude
\d**=#1\Lengthunit\ifdim\d**<0pt\d**-\d**\fi
\multiply\N*4\divide\d**\N*\sm*\n*\*one\loop
\shl**{\dd*}\dt*=1.3\dd*\advance\n*\*one
\shl**{\dt*}\advance\n*\*one
\shl**{\dd*}\advance\n*\*two
\dd*-\dd*\ifnum\n*<\N*\repeat\n*\N*\shl**{0pt}}}

\def\w*lin(#1,#2){\rlap{\toks0={#1}\toks1={#2}\d**=\Lengthunit\dd*=-.12\d**
\divide\dd*\*ths \multiply\dd*\magnitude
\N*8\divide\d**\N*\sm*\n*\*one\loop
\shl**{\dd*}\dt*=1.3\dd*\advance\n*\*one
\shl**{\dt*}\advance\n*\*one
\shl**{\dd*}\advance\n*\*one
\shl**{0pt}\dd*-\dd*\advance\n*1\ifnum\n*<\N*\repeat}}

\def\l*arc(#1,#2)[#3][#4]{\rlap{\toks0={#1}\toks1={#2}\d**=\Lengthunit
\dd*=#3.037\d**\dd*=#4\dd*\dt*=#3.049\d**\dt*=#4\dt*\ifdim\d**>10mm\relax
\d**=.25\d**\n*\*one\shl**{-\dd*}\n*\*two\shl**{-\dt*}\n*3\relax
\shl**{-\dd*}\n*4\relax\shl**{0pt}\else
\ifdim\d**>5mm\d**=.5\d**\n*\*one\shl**{-\dt*}\n*\*two
\shl**{0pt}\else\n*\*one\shl**{0pt}\fi\fi}}

\def\d*arc(#1,#2)[#3][#4]{\rlap{\toks0={#1}\toks1={#2}\d**=\Lengthunit
\dd*=#3.037\d**\dd*=#4\dd*\d**=.25\d**\sm*\n*\*one\shl**{-\dd*}\relax
\n*3\relax\sh*(#1,#2){\xL*=\xscale\dd*\yL*=\yscale\dd*
\kern#2\xL*\lower#1\yL*\hbox{\sm*}}\n*4\relax\shl**{0pt}}}

\def\shl**#1{\c*=\the\n*\d**\d*=#1\relax
\a*=\the\toks0\c*\b*=\the\toks1\d*\advance\a*-\b*
\b*=\the\toks1\c*\d*=\the\toks0\d*\advance\b*\d*
\a*=\xscale\a*\b*=\yscale\b*
\rx* \the\cos*\a* \tmp* \the\sin*\b* \advance\rx*-\tmp*
\ry* \the\cos*\b* \tmp* \the\sin*\a* \advance\ry*\tmp*
\raise\ry*\rlap{\kern\rx*\unhcopy\spl*}}

\def\wlin*#1(#2,#3)[#4]{\rlap{\toks0={#2}\toks1={#3}\relax
\c*=#1\l*\c*\c*=.01\Lengthunit\m*\c*\divide\l*\m*
\c*=\the\Nhalfperiods5sp\multiply\c*\l*\N*\c*\divide\N*\*ths
\divide\N*2\multiply\N*2\advance\N*\*one
\dd*=.002\Lengthunit\dd*=#4\dd*\multiply\dd*\l*\divide\dd*\N*
\divide\dd*\*ths \multiply\dd*\magnitude
\d**=#1\multiply\N*4\divide\d**\N*\sm*\n*\*one\loop
\shl**{\dd*}\dt*=1.3\dd*\advance\n*\*one
\shl**{\dt*}\advance\n*\*one
\shl**{\dd*}\advance\n*\*two
\dd*-\dd*\ifnum\n*<\N*\repeat\n*\N*\shl**{0pt}}}

\def\wavebox#1{\setbox0\hbox{#1}\relax
\a*=\wd0\advance\a*14pt\b*=\ht0\advance\b*\dp0\advance\b*14pt\relax
\hbox{\kern9pt\relax
\rmov*(0pt,\ht0){\rmov*(-7pt,7pt){\wlin*\a*(1,0)[+]\wlin*\b*(0,-1)[-]}}\relax
\rmov*(\wd0,-\dp0){\rmov*(7pt,-7pt){\wlin*\a*(-1,0)[+]\wlin*\b*(0,1)[-]}}\relax
\box0\kern9pt}}

\def\rectangle#1(#2,#3){\relax
\lin#1(#2,0)\lin#1(0,#3)\mov#1(0,#3){\lin#1(#2,0)}\mov#1(#2,0){\lin#1(0,#3)}}

\def\dashrectangle#1(#2,#3){\dashlin#1(#2,0)\dashlin#1(0,#3)\relax
\mov#1(0,#3){\dashlin#1(#2,0)}\mov#1(#2,0){\dashlin#1(0,#3)}}

\def\waverectangle#1(#2,#3){\L*=#1\Lengthunit\a*=#2\L*\b*=#3\L*
\ifdim\a*<0pt\a*-\a*\def\x*{-1}\else\def\x*{1}\fi
\ifdim\b*<0pt\b*-\b*\def\y*{-1}\else\def\y*{1}\fi
\wlin*\a*(\x*,0)[-]\wlin*\b*(0,\y*)[+]\relax
\mov#1(0,#3){\wlin*\a*(\x*,0)[+]}\mov#1(#2,0){\wlin*\b*(0,\y*)[-]}}

\def\calcparab*{\ifnum\n*>\m*\k*\N*\advance\k*-\n*\else\k*\n*\fi
\a*=\the\k* sp\a*=10\a*\b*\dm*\advance\b*-\a*\k*\b*
\a*=\the\*ths\b*\divide\a*\l*\multiply\a*\k*
\divide\a*\l*\k*\*ths\r*\a*\advance\k*-\r*\dt*=\the\k*\L*}

\def\arcto#1(#2,#3)[#4]{\rlap{\toks0={#2}\toks1={#3}\calcnum*#1(#2,#3)\relax
\dm*=135sp\dm*=#1\dm*\d**=#1\Lengthunit\ifdim\dm*<0pt\dm*-\dm*\fi
\multiply\dm*\r*\a*=.3\dm*\a*=#4\a*\ifdim\a*<0pt\a*-\a*\fi
\advance\dm*\a*\N*\dm*\divide\N*10000\relax
\divide\N*2\multiply\N*2\advance\N*\*one
\L*=-.25\d**\L*=#4\L*\divide\d**\N*\divide\L*\*ths
\m*\N*\divide\m*2\dm*=\the\m*5sp\l*\dm*\sm*\n*\*one\loop
\calcparab*\shl**{-\dt*}\advance\n*1\ifnum\n*<\N*\repeat}}

\def\arrarcto#1(#2,#3)[#4]{\L*=#1\Lengthunit\L*=.54\L*
\arcto#1(#2,#3)[#4]\rmov*(#2\L*,#3\L*){\d*=.457\L*\d*=#4\d*\d**-\d*
\rmov*(#3\d**,#2\d*){\arrow.02(#2,#3)}}}

\def\dasharcto#1(#2,#3)[#4]{\rlap{\toks0={#2}\toks1={#3}\relax
\calcnum*#1(#2,#3)\dm*=\the\N*5sp\a*=.3\dm*\a*=#4\a*\ifdim\a*<0pt\a*-\a*\fi
\advance\dm*\a*\N*\dm*
\divide\N*20\multiply\N*2\advance\N*1\d**=#1\Lengthunit
\L*=-.25\d**\L*=#4\L*\divide\d**\N*\divide\L*\*ths
\m*\N*\divide\m*2\dm*=\the\m*5sp\l*\dm*
\sm*\n*\*one\loop\calcparab*
\shl**{-\dt*}\advance\n*1\ifnum\n*>\N*\else\calcparab*
\sh*(#2,#3){\xL*=#3\dt* \yL*=#2\dt*
\rx* \the\cos*\xL* \tmp* \the\sin*\yL* \advance\rx*\tmp*
\ry* \the\cos*\yL* \tmp* \the\sin*\xL* \advance\ry*-\tmp*
\kern\rx*\lower\ry*\hbox{\sm*}}\fi
\advance\n*1\ifnum\n*<\N*\repeat}}

\def\*shl*#1{\c*=\the\n*\d**\advance\c*#1\a**\d*\dt*\advance\d*#1\b**
\a*=\the\toks0\c*\b*=\the\toks1\d*\advance\a*-\b*
\b*=\the\toks1\c*\d*=\the\toks0\d*\advance\b*\d*
\rx* \the\cos*\a* \tmp* \the\sin*\b* \advance\rx*-\tmp*
\ry* \the\cos*\b* \tmp* \the\sin*\a* \advance\ry*\tmp*
\raise\ry*\rlap{\kern\rx*\unhcopy\spl*}}

\def\calcnormal*#1{\b**=10000sp\a**\b**\k*\n*\advance\k*-\m*
\multiply\a**\k*\divide\a**\m*\a**=#1\a**\ifdim\a**<0pt\a**-\a**\fi
\ifdim\a**>\b**\d*=.96\a**\advance\d*.4\b**
\else\d*=.96\b**\advance\d*.4\a**\fi
\d*=.01\d*\r*\d*\divide\a**\r*\divide\b**\r*
\ifnum\k*<0\a**-\a**\fi\d*=#1\d*\ifdim\d*<0pt\b**-\b**\fi
\k*\a**\a**=\the\k*\dd*\k*\b**\b**=\the\k*\dd*}

\def\wavearcto#1(#2,#3)[#4]{\rlap{\toks0={#2}\toks1={#3}\relax
\calcnum*#1(#2,#3)\c*=\the\N*5sp\a*=.4\c*\a*=#4\a*\ifdim\a*<0pt\a*-\a*\fi
\advance\c*\a*\N*\c*\divide\N*20\multiply\N*2\advance\N*-1\multiply\N*4\relax
\d**=#1\Lengthunit\dd*=.012\d**
\divide\dd*\*ths \multiply\dd*\magnitude
\ifdim\d**<0pt\d**-\d**\fi\L*=.25\d**
\divide\d**\N*\divide\dd*\N*\L*=#4\L*\divide\L*\*ths
\m*\N*\divide\m*2\dm*=\the\m*0sp\l*\dm*
\sm*\n*\*one\loop\calcnormal*{#4}\calcparab*
\*shl*{1}\advance\n*\*one\calcparab*
\*shl*{1.3}\advance\n*\*one\calcparab*
\*shl*{1}\advance\n*2\dd*-\dd*\ifnum\n*<\N*\repeat\n*\N*\shl**{0pt}}}

\def\triangarcto#1(#2,#3)[#4]{\rlap{\toks0={#2}\toks1={#3}\relax
\calcnum*#1(#2,#3)\c*=\the\N*5sp\a*=.4\c*\a*=#4\a*\ifdim\a*<0pt\a*-\a*\fi
\advance\c*\a*\N*\c*\divide\N*20\multiply\N*2\advance\N*-1\multiply\N*2\relax
\d**=#1\Lengthunit\dd*=.012\d**
\divide\dd*\*ths \multiply\dd*\magnitude
\ifdim\d**<0pt\d**-\d**\fi\L*=.25\d**
\divide\d**\N*\divide\dd*\N*\L*=#4\L*\divide\L*\*ths
\m*\N*\divide\m*2\dm*=\the\m*0sp\l*\dm*
\sm*\n*\*one\loop\calcnormal*{#4}\calcparab*
\*shl*{1}\advance\n*2\dd*-\dd*\ifnum\n*<\N*\repeat\n*\N*\shl**{0pt}}}

\def\hr*#1{\L*=\xscale\Lengthunit\ifnum
\angle**=0\clap{\vrule width#1\L* height.1pt}\else
\L*=#1\L*\L*=.5\L*\rmov*(-\L*,0pt){\sm*}\rmov*(\L*,0pt){\sl*}\fi}

\def\shade#1[#2]{\rlap{\Lengthunit=#1\Lengthunit
\special{em:linewidth .001pt}\relax
\mov(0,#2.05){\hr*{.994}}\mov(0,#2.1){\hr*{.980}}\relax
\mov(0,#2.15){\hr*{.953}}\mov(0,#2.2){\hr*{.916}}\relax
\mov(0,#2.25){\hr*{.867}}\mov(0,#2.3){\hr*{.798}}\relax
\mov(0,#2.35){\hr*{.715}}\mov(0,#2.4){\hr*{.603}}\relax
\mov(0,#2.45){\hr*{.435}}\special{em:linewidth \the\linwid*}}}

\def\dshade#1[#2]{\rlap{\special{em:linewidth .001pt}\relax
\Lengthunit=#1\Lengthunit\if#2-\def\t*{+}\else\def\t*{-}\fi
\mov(0,\t*.025){\relax
\mov(0,#2.05){\hr*{.995}}\mov(0,#2.1){\hr*{.988}}\relax
\mov(0,#2.15){\hr*{.969}}\mov(0,#2.2){\hr*{.937}}\relax
\mov(0,#2.25){\hr*{.893}}\mov(0,#2.3){\hr*{.836}}\relax
\mov(0,#2.35){\hr*{.760}}\mov(0,#2.4){\hr*{.662}}\relax
\mov(0,#2.45){\hr*{.531}}\mov(0,#2.5){\hr*{.320}}\relax
\special{em:linewidth \the\linwid*}}}}

\def\vdot{\rlap{\kern-1.9pt\lower1.8pt\hbox{$\scriptstyle\bullet$}}}
\def\vtimes{\rlap{\kern-3pt\lower1.8pt\hbox{$\scriptstyle\times$}}}
\def\vDot{\rlap{\kern-2.3pt\lower2.7pt\hbox{$\bullet$}}}
\def\vTimes{\rlap{\kern-3.6pt\lower2.4pt\hbox{$\times$}}}

\def\arc(#1)[#2,#3]{{\k*=#2\l*=#3\m*=\l*
\advance\m*-6\ifnum\k*>\l*\relax\else
{\rotate(#2)\mov(#1,0){\sm*}}\loop
\ifnum\k*<\m*\advance\k*5{\rotate(\k*)\mov(#1,0){\sl*}}\repeat
{\rotate(#3)\mov(#1,0){\sl*}}\fi}}

\def\dasharc(#1)[#2,#3]{{\k**=#2\n*=#3\advance\n*-1\advance\n*-\k**
\L*=1000sp\L*#1\L* \multiply\L*\n* \multiply\L*\Nhalfperiods
\divide\L*57\N*\L* \divide\N*2000\ifnum\N*=0\N*1\fi
\r*\n*  \divide\r*\N* \ifnum\r*<2\r*2\fi
\m**\r* \divide\m**2 \l**\r* \advance\l**-\m** \N*\n* \divide\N*\r*
\k**\r* \multiply\k**\N* \dn*\n* 
\advance\dn*-\k** \divide\dn*2\advance\dn*\*one
\r*\l** \divide\r*2\advance\dn*\r* \advance\N*-2\k**#2\relax
\ifnum\l**<6{\rotate(#2)\mov(#1,0){\sm*}}\advance\k**\dn*
{\rotate(\k**)\mov(#1,0){\sl*}}\advance\k**\m**
{\rotate(\k**)\mov(#1,0){\sm*}}\loop
\advance\k**\l**{\rotate(\k**)\mov(#1,0){\sl*}}\advance\k**\m**
{\rotate(\k**)\mov(#1,0){\sm*}}\advance\N*-1\ifnum\N*>0\repeat
{\rotate(#3)\mov(#1,0){\sl*}}\else\advance\k**\dn*
\arc(#1)[#2,\k**]\loop\advance\k**\m** \r*\k**
\advance\k**\l** {\arc(#1)[\r*,\k**]}\relax
\advance\N*-1\ifnum\N*>0\repeat
\advance\k**\m**\arc(#1)[\k**,#3]\fi}}

\def\triangarc#1(#2)[#3,#4]{{\k**=#3\n*=#4\advance\n*-\k**
\L*=1000sp\L*#2\L* \multiply\L*\n* \multiply\L*\Nhalfperiods
\divide\L*57\N*\L* \divide\N*1000\ifnum\N*=0\N*1\fi
\d**=#2\Lengthunit \d*\d** \divide\d*57\multiply\d*\n*
\r*\n*  \divide\r*\N* \ifnum\r*<2\r*2\fi
\m**\r* \divide\m**2 \l**\r* \advance\l**-\m** \N*\n* \divide\N*\r*
\dt*\d* \divide\dt*\N* \dt*.5\dt* \dt*#1\dt*
\divide\dt*1000\multiply\dt*\magnitude
\k**\r* \multiply\k**\N* \dn*\n* \advance\dn*-\k** \divide\dn*2\relax
\r*\l** \divide\r*2\advance\dn*\r* \advance\N*-1\k**#3\relax
{\rotate(#3)\mov(#2,0){\sm*}}\advance\k**\dn*
{\rotate(\k**)\mov(#2,0){\sl*}}\advance\k**-\m**\advance\l**\m**\loop\dt*-\dt*
\d*\d** \advance\d*\dt*
\advance\k**\l**{\rotate(\k**)\rmov*(\d*,0pt){\sl*}}%
\advance\N*-1\ifnum\N*>0\repeat\advance\k**\m**
{\rotate(\k**)\mov(#2,0){\sl*}}{\rotate(#4)\mov(#2,0){\sl*}}}}

\def\wavearc#1(#2)[#3,#4]{{\k**=#3\n*=#4\advance\n*-\k**
\L*=4000sp\L*#2\L* \multiply\L*\n* \multiply\L*\Nhalfperiods
\divide\L*57\N*\L* \divide\N*1000\ifnum\N*=0\N*1\fi
\d**=#2\Lengthunit \d*\d** \divide\d*57\multiply\d*\n*
\r*\n*  \divide\r*\N* \ifnum\r*=0\r*1\fi
\m**\r* \divide\m**2 \l**\r* \advance\l**-\m** \N*\n* \divide\N*\r*
\dt*\d* \divide\dt*\N* \dt*.7\dt* \dt*#1\dt*
\divide\dt*1000\multiply\dt*\magnitude
\k**\r* \multiply\k**\N* \dn*\n* \advance\dn*-\k** \divide\dn*2\relax
\divide\N*4\advance\N*-1\k**#3\relax
{\rotate(#3)\mov(#2,0){\sm*}}\advance\k**\dn*
{\rotate(\k**)\mov(#2,0){\sl*}}\advance\k**-\m**\advance\l**\m**\loop\dt*-\dt*
\d*\d** \advance\d*\dt* \dd*\d** \advance\dd*1.3\dt*
\advance\k**\r*{\rotate(\k**)\rmov*(\d*,0pt){\sl*}}\relax
\advance\k**\r*{\rotate(\k**)\rmov*(\dd*,0pt){\sl*}}\relax
\advance\k**\r*{\rotate(\k**)\rmov*(\d*,0pt){\sl*}}\relax
\advance\k**\r*
\advance\N*-1\ifnum\N*>0\repeat\advance\k**\m**
{\rotate(\k**)\mov(#2,0){\sl*}}{\rotate(#4)\mov(#2,0){\sl*}}}}

\def\gmov*#1(#2,#3)#4{\rlap{\L*=#1\Lengthunit
\xL*=#2\L* \yL*=#3\L*
\rx* \gcos*\xL* \tmp* \gsin*\yL* \advance\rx*-\tmp*
\ry* \gcos*\yL* \tmp* \gsin*\xL* \advance\ry*\tmp*
\rx*=\xscale\rx* \ry*=\yscale\ry*
\xL* \the\cos*\rx* \tmp* \the\sin*\ry* \advance\xL*-\tmp*
\yL* \the\cos*\ry* \tmp* \the\sin*\rx* \advance\yL*\tmp*
\kern\xL*\raise\yL*\hbox{#4}}}

\def\rgmov*(#1,#2)#3{\rlap{\xL*#1\yL*#2\relax
\rx* \gcos*\xL* \tmp* \gsin*\yL* \advance\rx*-\tmp*
\ry* \gcos*\yL* \tmp* \gsin*\xL* \advance\ry*\tmp*
\rx*=\xscale\rx* \ry*=\yscale\ry*
\xL* \the\cos*\rx* \tmp* \the\sin*\ry* \advance\xL*-\tmp*
\yL* \the\cos*\ry* \tmp* \the\sin*\rx* \advance\yL*\tmp*
\kern\xL*\raise\yL*\hbox{#3}}}

\def\Earc(#1)[#2,#3][#4,#5]{{\k*=#2\l*=#3\m*=\l*
\advance\m*-6\ifnum\k*>\l*\relax\else\def\xscale{#4}\def\yscale{#5}\relax
{\angle**0\rotate(#2)}\gmov*(#1,0){\sm*}\loop
\ifnum\k*<\m*\advance\k*5\relax
{\angle**0\rotate(\k*)}\gmov*(#1,0){\sl*}\repeat
{\angle**0\rotate(#3)}\gmov*(#1,0){\sl*}\relax
\def\xscale{1}\def\yscale{1}\fi}}

\def\dashEarc(#1)[#2,#3][#4,#5]{{\k**=#2\n*=#3\advance\n*-1\advance\n*-\k**
\L*=1000sp\L*#1\L* \multiply\L*\n* \multiply\L*\Nhalfperiods
\divide\L*57\N*\L* \divide\N*2000\ifnum\N*=0\N*1\fi
\r*\n*  \divide\r*\N* \ifnum\r*<2\r*2\fi
\m**\r* \divide\m**2 \l**\r* \advance\l**-\m** \N*\n* \divide\N*\r*
\k**\r*\multiply\k**\N* \dn*\n* \advance\dn*-\k** \divide\dn*2\advance\dn*\*one
\r*\l** \divide\r*2\advance\dn*\r* \advance\N*-2\k**#2\relax
\ifnum\l**<6\def\xscale{#4}\def\yscale{#5}\relax
{\angle**0\rotate(#2)}\gmov*(#1,0){\sm*}\advance\k**\dn*
{\angle**0\rotate(\k**)}\gmov*(#1,0){\sl*}\advance\k**\m**
{\angle**0\rotate(\k**)}\gmov*(#1,0){\sm*}\loop
\advance\k**\l**{\angle**0\rotate(\k**)}\gmov*(#1,0){\sl*}\advance\k**\m**
{\angle**0\rotate(\k**)}\gmov*(#1,0){\sm*}\advance\N*-1\ifnum\N*>0\repeat
{\angle**0\rotate(#3)}\gmov*(#1,0){\sl*}\def\xscale{1}\def\yscale{1}\else
\advance\k**\dn* \Earc(#1)[#2,\k**][#4,#5]\loop\advance\k**\m** \r*\k**
\advance\k**\l** {\Earc(#1)[\r*,\k**][#4,#5]}\relax
\advance\N*-1\ifnum\N*>0\repeat
\advance\k**\m**\Earc(#1)[\k**,#3][#4,#5]\fi}}

\def\triangEarc#1(#2)[#3,#4][#5,#6]{{\k**=#3\n*=#4\advance\n*-\k**
\L*=1000sp\L*#2\L* \multiply\L*\n* \multiply\L*\Nhalfperiods
\divide\L*57\N*\L* \divide\N*1000\ifnum\N*=0\N*1\fi
\d**=#2\Lengthunit \d*\d** \divide\d*57\multiply\d*\n*
\r*\n*  \divide\r*\N* \ifnum\r*<2\r*2\fi
\m**\r* \divide\m**2 \l**\r* \advance\l**-\m** \N*\n* \divide\N*\r*
\dt*\d* \divide\dt*\N* \dt*.5\dt* \dt*#1\dt*
\divide\dt*1000\multiply\dt*\magnitude
\k**\r* \multiply\k**\N* \dn*\n* \advance\dn*-\k** \divide\dn*2\relax
\r*\l** \divide\r*2\advance\dn*\r* \advance\N*-1\k**#3\relax
\def\xscale{#5}\def\yscale{#6}\relax
{\angle**0\rotate(#3)}\gmov*(#2,0){\sm*}\advance\k**\dn*
{\angle**0\rotate(\k**)}\gmov*(#2,0){\sl*}\advance\k**-\m**
\advance\l**\m**\loop\dt*-\dt* \d*\d** \advance\d*\dt*
\advance\k**\l**{\angle**0\rotate(\k**)}\rgmov*(\d*,0pt){\sl*}\relax
\advance\N*-1\ifnum\N*>0\repeat\advance\k**\m**
{\angle**0\rotate(\k**)}\gmov*(#2,0){\sl*}\relax
{\angle**0\rotate(#4)}\gmov*(#2,0){\sl*}\def\xscale{1}\def\yscale{1}}}

\def\waveEarc#1(#2)[#3,#4][#5,#6]{{\k**=#3\n*=#4\advance\n*-\k**
\L*=4000sp\L*#2\L* \multiply\L*\n* \multiply\L*\Nhalfperiods
\divide\L*57\N*\L* \divide\N*1000\ifnum\N*=0\N*1\fi
\d**=#2\Lengthunit \d*\d** \divide\d*57\multiply\d*\n*
\r*\n*  \divide\r*\N* \ifnum\r*=0\r*1\fi
\m**\r* \divide\m**2 \l**\r* \advance\l**-\m** \N*\n* \divide\N*\r*
\dt*\d* \divide\dt*\N* \dt*.7\dt* \dt*#1\dt*
\divide\dt*1000\multiply\dt*\magnitude
\k**\r* \multiply\k**\N* \dn*\n* \advance\dn*-\k** \divide\dn*2\relax
\divide\N*4\advance\N*-1\k**#3\def\xscale{#5}\def\yscale{#6}\relax
{\angle**0\rotate(#3)}\gmov*(#2,0){\sm*}\advance\k**\dn*
{\angle**0\rotate(\k**)}\gmov*(#2,0){\sl*}\advance\k**-\m**
\advance\l**\m**\loop\dt*-\dt*
\d*\d** \advance\d*\dt* \dd*\d** \advance\dd*1.3\dt*
\advance\k**\r*{\angle**0\rotate(\k**)}\rgmov*(\d*,0pt){\sl*}\relax
\advance\k**\r*{\angle**0\rotate(\k**)}\rgmov*(\dd*,0pt){\sl*}\relax
\advance\k**\r*{\angle**0\rotate(\k**)}\rgmov*(\d*,0pt){\sl*}\relax
\advance\k**\r*
\advance\N*-1\ifnum\N*>0\repeat\advance\k**\m**
{\angle**0\rotate(\k**)}\gmov*(#2,0){\sl*}\relax
{\angle**0\rotate(#4)}\gmov*(#2,0){\sl*}\def\xscale{1}\def\yscale{1}}}

\newcount\CatcodeOfAtSign
\CatcodeOfAtSign=\the\catcode`\@
\catcode`\@=11
\def\@arc#1[#2][#3]{\rlap{\Lengthunit=#1\Lengthunit
\sm*\l*arc(#2.1914,#3.0381)[#2][#3]\relax
\mov(#2.1914,#3.0381){\l*arc(#2.1622,#3.1084)[#2][#3]}\relax
\mov(#2.3536,#3.1465){\l*arc(#2.1084,#3.1622)[#2][#3]}\relax
\mov(#2.4619,#3.3086){\l*arc(#2.0381,#3.1914)[#2][#3]}}}

\def\dash@arc#1[#2][#3]{\rlap{\Lengthunit=#1\Lengthunit
\d*arc(#2.1914,#3.0381)[#2][#3]\relax
\mov(#2.1914,#3.0381){\d*arc(#2.1622,#3.1084)[#2][#3]}\relax
\mov(#2.3536,#3.1465){\d*arc(#2.1084,#3.1622)[#2][#3]}\relax
\mov(#2.4619,#3.3086){\d*arc(#2.0381,#3.1914)[#2][#3]}}}

\def\wave@arc#1[#2][#3]{\rlap{\Lengthunit=#1\Lengthunit
\w*lin(#2.1914,#3.0381)\relax
\mov(#2.1914,#3.0381){\w*lin(#2.1622,#3.1084)}\relax
\mov(#2.3536,#3.1465){\w*lin(#2.1084,#3.1622)}\relax
\mov(#2.4619,#3.3086){\w*lin(#2.0381,#3.1914)}}}

\def\bezier#1(#2,#3)(#4,#5)(#6,#7){\N*#1\l*\N* \advance\l*\*one
\d* #4\Lengthunit \advance\d* -#2\Lengthunit \multiply\d* \*two
\b* #6\Lengthunit \advance\b* -#2\Lengthunit
\advance\b*-\d* \divide\b*\N*
\d** #5\Lengthunit \advance\d** -#3\Lengthunit \multiply\d** \*two
\b** #7\Lengthunit \advance\b** -#3\Lengthunit
\advance\b** -\d** \divide\b**\N*
\mov(#2,#3){\sm*{\loop\ifnum\m*<\l*
\a*\m*\b* \advance\a*\d* \divide\a*\N* \multiply\a*\m*
\a**\m*\b** \advance\a**\d** \divide\a**\N* \multiply\a**\m*
\rmov*(\a*,\a**){\unhcopy\spl*}\advance\m*\*one\repeat}}}

\catcode`\*=12

\newcount\n@ast

\def\n@ast@#1{\n@ast0\relax\get@ast@#1\end}
\def\get@ast@#1{\ifx#1\end\let\next\relax\else
\ifx#1*\advance\n@ast1\fi\let\next\get@ast@\fi\next}

\newif\if@up \newif\if@dwn
\def\up@down@#1{\@upfalse\@dwnfalse
\if#1u\@uptrue\fi\if#1U\@uptrue\fi\if#1+\@uptrue\fi
\if#1d\@dwntrue\fi\if#1D\@dwntrue\fi\if#1-\@dwntrue\fi}

\def\halfcirc#1(#2)[#3]{{\Lengthunit=#2\Lengthunit\up@down@{#3}\relax
\if@up\mov(0,.5){\@arc[-][-]\@arc[+][-]}\fi
\if@dwn\mov(0,-.5){\@arc[-][+]\@arc[+][+]}\fi
\def\lft{\mov(0,.5){\@arc[-][-]}\mov(0,-.5){\@arc[-][+]}}\relax
\def\rght{\mov(0,.5){\@arc[+][-]}\mov(0,-.5){\@arc[+][+]}}\relax
\if#3l\lft\fi\if#3L\lft\fi\if#3r\rght\fi\if#3R\rght\fi
\n@ast@{#1}\relax
\ifnum\n@ast>0\if@up\shade[+]\fi\if@dwn\shade[-]\fi\fi
\ifnum\n@ast>1\if@up\dshade[+]\fi\if@dwn\dshade[-]\fi\fi}}

\def\halfdashcirc(#1)[#2]{{\Lengthunit=#1\Lengthunit\up@down@{#2}\relax
\if@up\mov(0,.5){\dash@arc[-][-]\dash@arc[+][-]}\fi
\if@dwn\mov(0,-.5){\dash@arc[-][+]\dash@arc[+][+]}\fi
\def\lft{\mov(0,.5){\dash@arc[-][-]}\mov(0,-.5){\dash@arc[-][+]}}\relax
\def\rght{\mov(0,.5){\dash@arc[+][-]}\mov(0,-.5){\dash@arc[+][+]}}\relax
\if#2l\lft\fi\if#2L\lft\fi\if#2r\rght\fi\if#2R\rght\fi}}

\def\halfwavecirc(#1)[#2]{{\Lengthunit=#1\Lengthunit\up@down@{#2}\relax
\if@up\mov(0,.5){\wave@arc[-][-]\wave@arc[+][-]}\fi
\if@dwn\mov(0,-.5){\wave@arc[-][+]\wave@arc[+][+]}\fi
\def\lft{\mov(0,.5){\wave@arc[-][-]}\mov(0,-.5){\wave@arc[-][+]}}\relax
\def\rght{\mov(0,.5){\wave@arc[+][-]}\mov(0,-.5){\wave@arc[+][+]}}\relax
\if#2l\lft\fi\if#2L\lft\fi\if#2r\rght\fi\if#2R\rght\fi}}

\catcode`\*=11

\def\Circle#1(#2){\halfcirc#1(#2)[u]\halfcirc#1(#2)[d]\n@ast@{#1}\relax
\ifnum\n@ast>0\L*=\xscale\Lengthunit
\ifnum\angle**=0\clap{\vrule width#2\L* height.1pt}\else
\L*=#2\L*\L*=.5\L*\special{em:linewidth .001pt}\relax
\rmov*(-\L*,0pt){\sm*}\rmov*(\L*,0pt){\sl*}\relax
\special{em:linewidth \the\linwid*}\fi\fi}

\catcode`\*=12

\def\wavecirc(#1){\halfwavecirc(#1)[u]\halfwavecirc(#1)[d]}
\def\dashcirc(#1){\halfdashcirc(#1)[u]\halfdashcirc(#1)[d]}

\def\xscale{1}

\def\yscale{1}

\def\Ellipse#1(#2)[#3,#4]{\def\xscale{#3}\def\yscale{#4}\relax
\Circle#1(#2)\def\xscale{1}\def\yscale{1}}

\def\dashEllipse(#1)[#2,#3]{\def\xscale{#2}\def\yscale{#3}\relax
\dashcirc(#1)\def\xscale{1}\def\yscale{1}}

\def\waveEllipse(#1)[#2,#3]{\def\xscale{#2}\def\yscale{#3}\relax
\wavecirc(#1)\def\xscale{1}\def\yscale{1}}

\def\halfEllipse#1(#2)[#3][#4,#5]{\def\xscale{#4}\def\yscale{#5}\relax
\halfcirc#1(#2)[#3]\def\xscale{1}\def\yscale{1}}

\def\halfdashEllipse(#1)[#2][#3,#4]{\def\xscale{#3}\def\yscale{#4}\relax
\halfdashcirc(#1)[#2]\def\xscale{1}\def\yscale{1}}

\def\halfwaveEllipse(#1)[#2][#3,#4]{\def\xscale{#3}\def\yscale{#4}\relax
\halfwavecirc(#1)[#2]\def\xscale{1}\def\yscale{1}}

\catcode`\@=\the\CatcodeOfAtSign


\begin{center}
{\Large\bf
Superfield Effective Action in the Noncommutative Wess-Zumino Model}
\end{center}

\begin{center}
{\sl I. L. Buchbinder$^{a}$, M. Gomes$^{b}$, 
A. Yu. Petrov$^{a,b}$ and V. O. Rivelles$^{b}$}

\footnotesize
\vspace{.5cm}
{\it
$^{a}$ Department of Theoretical Physics,\\
Tomsk State Pedagogical University\\
Tomsk 634041, Russia\\
E-mail: joseph, petrov@tspu.edu.ru}\\
\vspace{.3cm}
{\it
$^{b}$ Instituto de F\'\i sica,\\
Universidade de S\~{a}o Paulo,\\
C. Postal 66318, 05315-970, S\~{a}o Paulo, SP, Brazil\\
E-mail: mgomes, petrov, rivelles@fma.if.usp.br}\\

\end{center}

\vspace*{.2cm}

\normalsize

\begin{abstract}
We introduce the concept of superfield effective action in
noncommutative ${\cal N}=1$ supersymmetric field theories containing chiral
superfields. One and two loops low-energy contributions to the 
effective action are found for the noncommutative Wess-Zumino model.
The one loop K\"{a}hlerian effective potential coincides 
with its commutative counterpart.
We show that the two loops nonplanar contributions to the K\"{a}hlerian
effective potential are leading in the case of small
noncommutativity. The structure of the 
leading chiral corrections to the effective action and the behaviour of
the chiral effective Lagrangian in the limit of large noncommutativity
are also investigated.
\end{abstract}


Nowadays, an enormous effort is being done to understand the properties
of noncommutative field theories. There are two main reasons for
that. By one side, they are the field theory limit of open strings in
the presence of a constant $B$-field \cite{Seiberg-Witten}. On the
other side, although being nonlocal field theories, they are still
tractable giving rise to new and interesting phenomena
\cite{Minwalla}.  In particular, in spite of their nonlocality, 
noncommutative models allow the construction of causal quantum field
theories. 

The main characteristic of noncommutative field theories is the
mixture of ultraviolet and infrared divergences which may turn the
ordinary (commutative) renormalizable theories into nonrenormalizable
ones \cite{Arefeva}. Supersymmetry seems to be needed to recover
renormalizability at least for the case of non-gauge theories
\cite{saolourenco}. The
complex scalar field theory with interaction 
$\phi^* \star \phi^* \star \phi \star \phi$ is one loop
nonrenormalizable. However, its supersymmetric extension, the
noncommutative Wess-Zumino model is renormalizable to all loop orders 
\cite{Wess-Zumino}. In $2+1$ dimensions the dynamical mass generation
in the Gross-Neveu model is spoiled by noncommutativity. Also, the
noncommutative 
nonlinear sigma model turns out to be nonrenormalizable due to the
mixture of ultraviolet and infrared divergences. However, the
noncommutative supersymmetric nonlinear sigma model, which includes
both models above, is one loop renormalizable \cite{Sigma-Model}. For
supersymmetric gauge theories the situation is more involved since the
effective action has quantum corrections for nonplanar graphs which
require the introduction of generalized Moyal products 
\cite{gauge-theories}. 

An essential ingredient in quantum field theory is the effective
action. It allows the study of several aspects of quantum field
models including the structure of ultraviolet divergences, the 
infrared behaviour and  quantum symmetries. Therefore, the 
effective action is a valuable tool which will provide the necessary
means to investigate the problem of 
ultraviolet/infrared mixing in noncommutative supersymmetric 
theories and clarify how the noncommutativity can influence the 
known properties of standard supersymmetric field models.

In this paper we calculate the leading chiral correction to
the superfield effective action in the massless noncommutative
Wess-Zumino model. We consider the massless case because there are no
chiral corrections for the massive theory as in the commutative case
\cite{Buch4}. The first nonvanishing correction appears at the
two loops level, also as in the commutative case, 
and presents neither ultraviolet nor infrared divergences. We
also calculate the one and two loops contributions to the
K\"ahlerian effective potential. For the one loop case there is no dependence
on the noncommutativity parameter and the result coincides with the
commutative one. At two loops the K\"alerian effective potential has a
nonplanar part which strongly depends on the noncommutativity. 

The most natural way to study the effective action makes use of 
superspace concepts. The formulation of noncommutative supersymmetric field
theories in superspace has already been performed
\cite{superspace}. Noncommutativity is only introduced for 
bosonic coordinates, the Grassmannian coordinates still being taken as 
anticommuting (see nevertheless the attempts to construct a superspace with
non-anticommuting Grassmann coordinates \cite{Klemm}). 
In the commutative case the effective action in superspace was
developed in \cite{Buch2} (see also \cite{BK0}). Its
application to the low-energy leading contributions to the effective
action were found for several superfield theories
\cite{Buch4,Budc,Buch3,Buch5,McA}. In the noncommutative case, one
loop  quantum corrections to the effective action in superfield form
were investigated for the Wess-Zumino model \cite{Popp} and for gauge
theories \cite{Zanon}. However, a systematic  development of the
concept of superfield effective action still remains to be done in the
noncommutative case. So, this paper is also devoted
to carry out such a generalization for any order in perturbation
theory. We will obtain the noncommutative analogs of
\cite{Buch2,Buch3}, i.e., the K\"ahlerian and chiral effective
potentials for the noncommutative Wess-Zumino model.  

The noncommutative massless Wess-Zumino model in superspace has the
action 
\begin{equation}
\label{act}
S[\bar{\Phi},\Phi]=\int d^8 z \bar{\Phi}\Phi+
(\lambda\int d^6 z \Phi^{*3}+h.c.), 
\end{equation}
where $\Phi(z)$ and $\bar{\Phi}(z)$ are chiral and antichiral superfields
respectively, $\lambda$ is a real coupling constant and
$\Phi^{*3}=\Phi*\Phi*\Phi$. The interaction term has 
the following expression 
\begin{eqnarray}
\label{cube}
& &\int d^6 z \Phi^{*3}(z)=\int d^2\theta \int d^4 x \int d^4 k_1 d^4 k_2
d^4 k_3 e^{i(k_1+k_2+k_3)x}
e^{-i\sum_{i<j}^3 k_i\times k_j} \times
\nonumber\\
&\times&
\int d^4 x_1 d^4 x_2 d^4 x_3 e^{-k_1x_1-k_2x_2-k_3 x_3}
\Phi(x_1,\theta)\Phi(x_2,\theta)\Phi(x_3,\theta),
\end{eqnarray}
where $k_i \times k_j = k^\mu_i \theta_{\mu\nu} k^\nu_j$ and
$\theta_{\mu\nu}$ is the noncommutativity parameter. 
The propagator is (we use the conventions of \cite{BK0}) 
\begin{eqnarray}
\label{prop}
<\Phi(z_1)\bar{\Phi}(z_2) > =\frac{\bar{D}^2
  D^2}{16\Box}\delta^8(z_1-z_2), 
\end{eqnarray}
and has the same expression an in the commutative case. 
The vertex is, however, modified. It reads \cite{Popp}
\bea
\label{vert}
\lambda{(2\pi)}^4\delta(k+l+p)\cos(k\times l),
\eea
where $k,l,p$ are the momenta of the superfields associated to the
vertex. 

The effective action $\Gamma [\bar{\Phi},\Phi]$ can be presented as a series
in supercovariant derivatives
$D_A=(\partial_a,D_{\alpha},\bar{D}_{\dot{\alpha}})$ in the form 
\begin{eqnarray}
\label{eac}
 \Gamma[\bar{\Phi},\Phi] &=& \int d^8z {\cal L}_{eff}
(\Phi,D_A\Phi,D_A D_B\Phi,\ldots,\bar{\Phi},D_A\bar{\Phi},D_A
D_B\bar{\Phi},\ldots)
+\nonumber\\
&+&(\int d^6z {\cal
  L}^{(c)}_{eff}(\Phi,\partial_a\Phi,\partial_a\partial_b\Phi,\ldots)
+ h.c.),
\end{eqnarray}
where 
${\cal L}_{eff}$ is the general effective Lagrangian and 
${\cal L}^{(c)}_{eff}$ is the chiral effective Lagrangian. 
It is clear that these effective Lagrangians contain the effects 
induced by noncommutativity. Our purpose is to find the leading
low-energy contributions to the effective Lagrangians. We assume that
they have the structure 
\begin{eqnarray}
\label{ep1}
{\cal L}_{eff}&=&K_{eff}{(\bar{\Phi},\Phi)}_*+\ldots
=\bar{\Phi}\Phi+\sum_{n=1}^{\infty}K^{(n)}_{eff} {(\bar{\Phi},\Phi)} +
\ldots,\\ 
\label{ep2}
{\cal L}^{(c)}_{eff}&=&W_{eff}{(\Phi)}_* =
\lambda\Phi^{*3}+\sum_{n=1}^{\infty}W^{(n)}_{eff} {(\Phi)}+\ldots,
\end{eqnarray}
where dots in Eq.(\ref{ep1}) mean space-time derivatives of 
the superfields $\bar\Phi$ and $\Phi$, and dots in Eq.(\ref{ep2}) mean
space-time derivatives  of $\Phi$. From these derivative dependent
terms we will keep only the leading ones in momentum and in the
noncommutativity parameter $\theta_{\mu \nu}$. In Eq.(\ref{ep1}) we
call $K_{eff}(\bar \Phi, 
\Phi, ...)$ the K\"{a}hlerian effective potential in noncommutative
theory and  $W_{eff}(\Phi, ...)$, 
in Eq.(\ref{ep2}), the chiral (or holomorphic) effective potential. 
Here $K^{(n)}_{eff}(\Phi,\bar{\Phi})$ is the $n$-th
correction to the K\"ahlerian potential and $W^{(n)}_{eff}(\Phi)$ is the
$n$-th correction to the chiral potential. 

To consider further the effective Lagrangians ${\cal L}_{eff}$ and
${\cal L}^{(c)}_{eff}$ we use the path integral representation of
the effective action \cite{BK0,BO}
\begin{eqnarray}
\label{Green1}
&& \exp(\frac{i}{\hbar}\Gamma[\bar{\Phi},\Phi]) = \nonumber\\
&& \int {\cal D} \phi {\cal D} \bar{\phi}  \exp\big( \frac{i}{\hbar}
S[\bar{\Phi}+\sqrt{\hbar}\bar{\phi},\Phi+\sqrt{\hbar}\phi]  -
\frac{1}{\sqrt{\hbar}}(\int d^6 z
\frac{\delta\Gamma[\bar{\Phi},\Phi]}{\delta\Phi(z)}\phi(z)+h.c. )
\big),
\end{eqnarray}
where $\Phi$ and $\bar{\Phi}$ are the background superfields and
$\phi$ and $\bar{\phi}$ are the quantum ones. The effective action can be
written as
$\Gamma[\bar{\Phi},\Phi]=S[\bar{\Phi},\Phi]+\tilde{\Gamma}
[\bar{\Phi},\Phi]$, 
where $\tilde{\Gamma}[\bar{\Phi},\Phi]$ is the quantum correction to
the classical action. Then Eq.(\ref{Green1}) 
allows us to obtain $\tilde{\Gamma}[\bar{\Phi},\Phi]$ in the form of a
loop expansion
$\label{Gamma}
 \tilde{\Gamma}[\bar{\Phi},\Phi] = \sum_{n=1}^{\infty}\hbar^n
\Gamma^{(n)} [\bar{\Phi},\Phi]$
and hence we get the loop expansion for the effective Lagrangians
${\cal L}_{eff}$ and ${\cal L}^{(c)}_{eff}$.

To find the loop corrections $\Gamma^{(n)} [\bar{\Phi},\Phi]$ in explicit
form we expand the right-hand side of Eq.(\ref{Green1}) in a power
series in the quantum superfields $\phi$, $\bar{\phi}$. For slowly
varying background fields in space-time, the quadratic part 
of the expansion of
$\frac{1}{\hbar}S[\bar{\Phi}+\sqrt{\hbar}\bar{\phi},\Phi+\sqrt{\hbar}\phi]$
in quantum superfields $\phi,\bar{\phi}$ is given by
\begin{eqnarray}
\label{qua}
S_2=\frac{1}{2}\int d^8 z \left(\begin{array}{cc}\phi&\bar{\phi}
\end{array}\right)
\left(\begin{array}{cc}
\lambda\Phi&(-\frac{1}{4})D^2\\
(-\frac{1}{4})\bar{D}^2&\lambda\bar{\Phi}
\end{array}\right)
\left(\begin{array}{c}\phi\\
\bar{\phi}
\end{array}
\right).
\end{eqnarray}
No Moyal product is present because $\Phi$ is a slowly
varying superfield. It means that the full low-energy 
one loop effective action in the
noncommutative theory will be the same as in the corresponding commutative
one \cite{Buch1,GR,PW}. We can expect non-trivial corrections due to
the noncommutativity only in the two loops approximation.

To find the two loops correction to the K\"ahlerian potential we need
to calculate the superpropagator associated to Eq.(\ref{qua}). It is
given by the solution of 
\bea
\label{matprop}
\left(\begin{array}{cc}
\lambda\Phi&(-\frac{1}{4})D^2\\
(-\frac{1}{4})\bar{D}^2&\lambda\bar{\Phi}
\end{array}\right)
\left(\begin{array}{cc}
G_{++}&G_{+-}\\
G_{-+}&G_{--}
\end{array}\right) = 
- \left(\begin{array}{cc}
\delta_+&0\\
0&\delta_-
\end{array}\right),
\eea
where $\delta_+ = -\frac{1}{4} \bar{D}^2 \delta^8(z_1-z_2)$ and
$\delta_- = -\frac{1}{4} {D}^2 \delta^8(z_1-z_2)$.  
The components of the matrix superpropagator for the case of constant
superfields is given by  
\begin{eqnarray}
\label{sol1}
G_{++}&=&\frac{\lambda\bar{\Phi}}{\Box+\lambda^2{|\Phi|}^2}
\frac{\bar{D}^2_1}{4}\delta_{12},\ \
G_{+-}=\frac{1}{\Box+\lambda^2{|\Phi|}^2}
\frac{\bar{D}^2_1 D^2_2}{16}\delta_{12},\nonumber\\
G_{-+}&=&\frac{1}{\Box+\lambda^2{|\Phi|}^2}
\frac{D^2_1\bar{D}^2_2}{16}\delta_{12},\ \
G_{--}=\frac{\lambda\Phi}{\Box+\lambda^2{|\Phi|}^2}
\frac{D^2_1}{4}\delta_{12}.
\end{eqnarray}
Since the K\"ahlerian effective potential depends only on the
superfields $\Phi$ and $\bar{\Phi}$ but 
not on their derivatives, supergraphs contributing to 
it must include an equal number of $D^2$ and $\bar{D}^2$ factors with all
vertices rewritten in the form of an integral over the whole superspace.
The only supergraph with equal number of $D^2$ and $\bar{D}^2$ factors is
given by:

\hspace{5cm}
\unitlength=.7mm
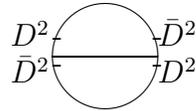
\begin{center}
\begin{picture}(12,12)
\put(0,0){\circle{20}}\put(-10,0){\line(1,0){20}}
\put(-18,2){$D^2$} \put(-18,-5){$\bar{D}^2$} \put(10,2){$\bar{D}^2$}
\put(10,-5){$D^2$}
\put(-10,2){-} \put(-10,-3){-} \put(9,2){-} \put(9,-3){-}
\put(2,-12){}
\end{picture}
\end{center}
\phantom{Two loops supergraph}

\vspace*{3mm}

\noindent 
The contribution of the supergraph, after evident D-algebra
manipulations, takes the form
\bea
\label{int0}
K^{(2)}=\frac{\lambda^2}{6}\int \frac{d^4kd^4l}{(2\pi)^8}
\cos^2(k\times l)\frac{1}{(k^2+m^2)(l^2+m^2)((k+l)^2+m^2)},
\eea
where $m^2\equiv \lambda^2|\Phi|^2$. 

This can be split into a planar and a nonplanar
part. The planar part is given by the integral
\bea
K^{(2)}_{pl}=\frac{\lambda^2}{12}\int \frac{d^4kd^4l}{(2\pi)^8}
\frac{1}{(k^2+m^2)(l^2+m^2)((k+l)^2+m^2)},
\eea
whose contribution is analogous to the contribution for the commutative 
Wess-Zumino model. The only difference is a multiplicative factor of 
$\frac{1}{2}$. After calculating the integrals and subtracting
divergences we get (cf. \cite{Buch4})  
\bea
K^{(2)}_{pl}=\frac{\lambda^2}{2(4\pi)^4}m^2
\Big(-\frac{1}{4}\log^2\frac{m^2}{\mu^2}
+\frac{3-\gamma}{2}\log\frac{m^2}{\mu^2}+\frac{3}{2}(\gamma-1)+\frac{1}{4}
(\gamma^2+\zeta(2))-b
\Big),
\eea
where $b$ is a finite constant whose origin is due to the choice of a
non-minimal subtraction scheme. 
Its value has to be fixed by proper normalization conditions.
We renormalized only the planar part
since, as we shall show, the nonplanar part is finite. 

To evaluate the nonplanar part of Eq.(\ref{int0}) let us consider the
integral 
\bea
\label{npi}
\int\frac{d^4k}{(2\pi)^4}\frac{e^{2i(k\times l)}}
{(k^2+m^2)((k+l)^2+m^2)}.
\eea
Using the Feynman representation and the $\alpha$-representation for
the denominator we can perform the integration over the momenta
arriving at 
\bea
\label{alfa}
\frac{1}{16\pi^2}\int_0^1 dx\int_0^{\infty}\frac{d\alpha}{\alpha}
e^{-\alpha(m^2+l^2x(1-x))-\frac{l\circ l}{\alpha}},
\eea
where $l\circ l\equiv l^a (\theta^2)_{ab} l^b$ and 
$(\theta^2)_{ab}=\theta_{ac}{\theta^c}_b$.
We note that if we set $\theta=0$ the integral 
becomes divergent due to the absence of the factor $e^{-l\circ l/\alpha}$.
Then, the nonplanar contribution to Eq.(\ref{int0}) is 
\bea
K^{(2)}_{np}=\frac{\lambda^2}{24}\frac{1}{16\pi^2}\int_0^1 dx\int_0^{\infty}
\frac{d\alpha}{\alpha}\int\frac{d^4l}{(2\pi)^4}\frac{1}{l^2+m^2}
e^{-\alpha(m^2+l^2x(1-x))-\frac{l\circ l}{\alpha}}.
\eea
Finally, we can exponentiate  $\frac{1}{l^2+m^2}$ to arrive at 
\bea
K^{(2)}_{np}=\frac{\lambda^2}{24}\frac{1}{16\pi^2} \int_0^1 dx \int_0^{\infty}
\frac{d\alpha}{\alpha} \int_0^\infty dz
e^{-(\alpha+z)m^2}\int\frac{d^4l}{(2\pi)^4} \exp[-l^m A_{mn} l^n],
\eea
where $A_{mn}$ is a matrix of the form
\bea
A_{mn}=\eta_{mn}(\alpha x(1-x)+z)+\frac{1}{\alpha}(\theta^2)_{mn}.
\eea
Notice that all integrals are convergent.
After a Wick rotation we can perform the integration over the 
momenta obtaining 
\bea
K^{(2)}_{np}=\frac{\lambda^2}{24}\frac{1}{(16\pi^2)^2} \int_0^1 dx
\int_0^{\infty} \frac{d\alpha}{\alpha} \int_0^\infty dz
e^{-(\alpha+z)m^2}{\det} ^{-1/2} [A_{mn}].
\eea

Carrying out the remaining integrations is quite
complicated. Therefore we specialize the matrix $\theta_{\mu\nu}$ to its
canonical form with diagonal blocks. Furthermore, to avoid troubles with
causality we allow only space-space noncommutativity. Then, the 
nonvanishing components are $\theta_{23}= - \theta_{32} =
a$, with $a$ having mass dimension $-2$. Hence, the nonvanishing
components of $\theta^2$ are $(\theta^2)_{22} =
(\theta^2)_{33} = - a^2$ and 
\bea
\label{det}
{\det} ^{-1/2} A=\frac{1}{(\alpha x(1-x)+z+\frac{a^2}{\alpha})
(\alpha x(1-x)+z)}.
\eea
Now we can rescale $\alpha m^2\to\alpha$ and $zm^2\to z$ so that the
new variables are dimensionless. We can also introduce a 
new dimensionless noncommutativity parameter $\tilde{a}^2=m^4a^2$.
Then the nonplanar correction takes the form
\bea
\label{npcor}
K^{(2)}_{np}=
m^2\frac{\lambda^2}{24}\frac{1}{(16\pi^2)^2}\int_0^1 dx\int_0^{\infty}
d\alpha \int_0^\infty dz 
\frac{e^{-(\alpha+z)}}{(\alpha^2x(1-x)+\alpha z+\tilde{a}^2)(\alpha
  x(1-x)+z)}. 
\eea
Eq.(\ref{npcor}) is the exact two loops nonplanar correction to the 
K\"{a}hlerian effective potential.

The integral in the right hand side of Eq.(\ref{npcor}) is still
complicated. However, there are two
limits of $\tilde{a}^2$ for which the integral can be performed.
Let us first consider the case $\tilde{a}^2>>1$. We can expand
Eq.(\ref{npcor}) in a power series in $\frac{1}{a}$ and arrive at
\bea
K^{(2)}_{np} &=&  m^2\frac{\lambda^2}{24}\frac{1}{(16\pi^2)^2}\int_0^1
dx\int_0^{\infty} d\alpha \int_0^\infty dz
\frac{\alpha}{\tilde{a}^2}\frac{e^{-(\alpha+z)}}{\alpha x(1-x)+z} 
(1-\frac{\alpha^2x(1-x)+\alpha z}{\tilde{a}^2}) + \nonumber \\ &+&
O(\frac{1}{\tilde{a}^6}).
\eea
After integration over $x,\alpha$ and $z$, and restoring the manifest
$\Phi$ dependence, the two loops nonplanar correction can be expressed
as 
\bea
K^{(2)}_{np}=
\frac{\lambda^4}{24}|\Phi|^2\frac{1}{(16\pi^2)^2}
(\frac{c_1}{\tilde{a}^2}+
\frac{c_2}{\tilde{a}^4})
+O(\frac{1}{\tilde{a}^6}),
\eea
where $c_1$ and $c_2$ are real numbers. Therefore the nonplanar
contribution is 
suppressed at large value of the noncommutativity parameter
$\tilde{a}$. Its leading term is proportional to
$\lambda^4|\Phi|^2\frac{1}{\tilde{a}^2}$.  Note that this
correction is finite and does not contain any singularity coming from
the UV/IR mixing. 

In the case of $\tilde{a}^2<<1$  we redefine the variables $\alpha$ and
$z$ by $\alpha'=\alpha a$ and $ z'=z a$, respectively.  As a result,
Eq.(\ref{npcor}) takes the form
\begin{equation}
\label{npcor1}
K^{(2)}_{np}=
m^2\frac{\lambda^2}{24}\frac{1}{a}\frac{1}{(16\pi^2)^2}
\int_0^1 dx\int_0^{\infty}
d\alpha' \int_0^\infty dz'
\frac{e^{-a(\alpha'+z')}}{({(\alpha')}^2x(1-x)+\alpha' z'+1)(\alpha'
  x(1-x)+z')},
\end{equation}
resulting in 
\bea
K^{(2)}_{np}=\frac{\lambda^4}{24}|\Phi|^2(\frac{d}{a}+O(a^0)).
\eea
where
\bea
d=\frac{1}{(16\pi^2)^2}
\int_0^1 dx\int_0^{\infty}
d\alpha' \int_0^\infty dz'
\frac{e^{-a(\alpha'+z')}}{({(\alpha')}^2x(1-x)+\alpha' z'+1)(\alpha'
  x(1-x)+z')},
\eea
is a constant.
In other words, in the case of small noncommutativity the nonplanar
correction becomes the leading one. This result agrees with the
predictions given in \cite{Huang} for the non-supersymmetric case.

Now let us turn to the evaluation of the corrections 
to the chiral effective potential. First we set the background
antichiral superfield $\bar{\Phi}$ to zero. Then the expansion of
$\frac{1}{\hbar}S[\bar{\Phi}+\sqrt{\hbar}\bar{\phi},\Phi+\sqrt{\hbar}\phi]$
in quantum superfields yields 
\begin{eqnarray}
\label{qchr}
S=\int d^8 z \phi\bar{\phi}+\lambda\int d^6 z
(3\Phi*\phi*\phi+\phi^{*3})+\lambda\int d^6 \bar{z}\bar{\phi}^{*3}.  
\end{eqnarray}
As we will show, chiral loop contributions begin at two
loops. Therefore we retain in Eq.(\ref{qchr}) only the terms of second
and third orders in quantum superfields (note that vertices of fourth order
in quantum superfields, which  in general are essential for
calculating the two loops effective action, are absent in this theory). 

The chiral action can be written as \cite{Popp} 
\bea
\label{star}
S_c=\int d^6 z \phi^{*3} =
\int d^2\q \int
\frac{d^4p_1 d^4p_2 }{(2\pi)^8}
e^{-p_1\times p_2}\phi(p_1,\q)\phi(p_2,\q)\phi(-(p_1+p_2),\q).
\eea
We see that the quantum $\phi^{*3}$ corrections have the same
structure as the original commutative interaction Lagrangian. The only 
difference is in the presence of an additional
factor $S(p_1,p_2)$ which arises after integration over internal
momenta. Then performing the inverse Fourier transformation we arrive
at the possible form for the quantum correction 
\bea
\label{qcr}
\Delta S_c&=&
\int d^2\q \int d^4 x_1 d^4 x_2 d^4 x_3
\frac{d^4p_1 d^4p_2 }{(2\pi)^8}
e^{-p_1\times p_2}\phi(x_1,\q)\phi(x_2,\q)\phi(x_3,\q)
\times\nonumber\\&\times&
e^{ix_1p_1+ix_2p_2+ix_3(-p_1-p_2)}
S(p_1,p_2).
\eea
We see that all quantum corrections are included in the single
function $S(p_1,p_2)$.
Assuming that the superfields under consideration are slowly varying
in space-time we can integrate over $x_2$ and $x_3$  and over the
momenta $k_1$ and $k_2$ getting  
\bea
\label{correc}
L_c=\int d^2\q\int d^4 x_1 \phi^3(x_1,\q)S(p_1,p_2)|_{p_1,p_2=0}.
\eea 
This correction has precisely the same form as that in the commutative
case. Thus, we showed that for slowly varying superfields their 
Moyal product coincides with their standard product.

The structure of the vertices and propagators are similar to those of
the commutative Wess-Zumino model and allow us to show
that there is only one supergraph contributing to the chiral effective
potential at two loops:

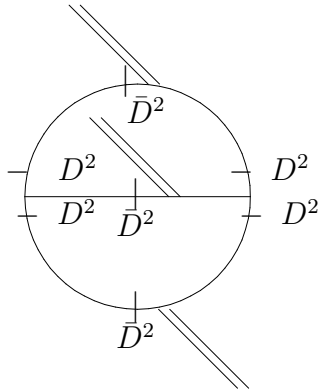
\begin{center}
\hspace{4cm}
\Lengthunit=1.5cm
\GRAPH(hsize=3){
\mov(.5,0){\Circle(2)\mov(-1,0){\lin(2,0)}
\ind(-2,10){|}
\ind(-2,-3){\bar{D}^2}\ind(-2,0){|}\ind(-2,-10){|}\ind(-2,-13){\bar{D}^2}
\ind(-2,7){\bar{D}^2}
\ind(-9,2){D^2} \ind(-10,-2){D^2} \ind(8,2){D^2} \ind(8,-2){D^2}
\ind(-18,2){-} \ind(-18,-2){-} \ind(0,2){-} \ind(0,-2){-}
\Linewidth{0.3pt}
\mov(-1,1){\lin(-.7,.7)}\mov(-1.1,1){\lin(-.7,.7)}
\mov(-1,-1){\lin(.7,-.7)}\mov(-1.1,-1){\lin(.7,-.7)}
\mov(-1,0){\lin(-.7,.7)}\mov(-1.1,0){\lin(-.7,.7)}}
}
\end{center}
\vspace*{3mm}

\noindent The double external lines denote the background
superfield $\Phi$. The superpropagator is given by
Eq.(\ref{prop}). The contribution of this supergraph is then 
\begin{eqnarray}
\label{I1}
&&\frac{\lambda^5}{12}
\int \frac{d^4p_1 d^4p_2}{{(2\pi)}^8}\frac{d^4k d^4l}{{(2\pi)}^8}
\int d^4\theta_1 d^4\theta_2 d^4\theta_3 d^4\theta_4 d^4\theta_5
\Phi(-p_1,\theta_3)
\Phi(-p_2,\theta_4)
\Phi(p_1+p_2,\theta_5) \times\nonumber\\&\times&
\frac{\cos(k\times l)\cos [(k+p_1)\times(l+p_2)] \cos(k\times p_1)
  \cos(l\times p_2)\cos[(k+l)\times(p_1+p_2)]}{k^2 l^2 {(k+p_1)}^2
  {(l+p_2)}^2 {(l+k)}^2 {(l+k+p_1+p_2)}^2} 
\times\nonumber\\&\times&
\delta_{13}\frac{\bar{D}^2_3}{4}\delta_{32}
\frac{D^2_1 \bar{D}^2_4}{16}\delta_{14}\delta_{42}
\frac{D^2_1 \bar{D}^2_5}{16}\delta_{15}\delta_{52}.
\end{eqnarray}
After $D$-algebra transformations, which can be carried out in the same
manner as in the commutative Wess-Zumino model, this expression can be
written as 
\begin{eqnarray}
\label{app}
&&\frac{\lambda^5}{12}
\int \frac{d^4p_1 d^4p_2}{{(2\pi)}^8}\frac{d^4k d^4l}{{(2\pi)}^8}
\int d^2\theta
\Phi(-p_1,\theta)
\Phi(-p_2,\theta)
\Phi(p_1+p_2,\theta)
\times\nonumber\\&\times&
\frac{k^2 p_1^2+ l^2 p_2^2 +2 (k l)(p_1 p_2)}
{k^2 l^2 {(k+p_1)}^2{(l+p_2)}^2{(l+k)}^2{(l+k+p_1+p_2)}^2}
\times\\&\times&
\cos(k\times l)\cos ((k+p_1)\times(l+p_2))
\cos(k\times p_1)\cos(l\times p_2)\cos((k+l)\times(p_1+p_2)),
\nonumber
\end{eqnarray}
where $kl = k^\mu l_\mu$. It has the same form as Eq.(\ref{qcr}), as
expected. We then find  
\bea
\label{s}
&&S(p_1,p_2)=\frac{\lambda^5}{12}
\int\frac{d^4 k d^4 l}{{(2\pi)}^8}\frac{k^2 p_2^2+ l^2
p_1^2 +2 (kl)(p_1 p_2)} {k^2 l^2
{(k+p_1)}^2{(l+p_2)}^2{(l+k)}^2{(l+k+p_1+p_2)}^2}
\times\\&\times& 
\cos(k\times l)\cos [(k+p_1)\times(l+p_2)]
\cos(k\times p_1)\cos(l\times p_2)
\cos[(k+l)\times(p_1+p_2)], \nonumber
\eea
where $p_1, p_2$ are regarded as external momenta. 
Therefore we need to analyze the behavior of $S(p_1,p_2)$  
Eq.(\ref{s}) in the limit $p_1,p_2\to 0$. It is natural to consider this
limit in the following way. We must first set one of these external
momenta (e.g. $p_2$) to zero, and then consider the limit of the 
expression as $p_1\to 0$. If we set $p_2=0$, multiply the cosine
factors and make several changes of variables we arrive at 
\bea
\label{sp}
&S(p)&=\frac{\lambda^5}{12}\frac{p^2}{8}
\int\frac{d^4 k d^4 l}{{(2\pi)}^8}
\Big[\frac{1}{k^2 {(k+p)}^2 l^2 {(l+k)}^2{(l+k+p)}^2}+\nonumber\\&+&
\frac{3\cos(2p\times l)}{k^2 {(k+p)}^2 l^2 {(l+k)}^2{(l+k+p)}^2} +
\frac{2\cos(2k\times l)}{k^2 {(k+p)}^2 l^2 {(l+k)}^2{(l+k+p)}^2}
\nonumber\\ 
&+&\frac{2\cos(2k\times l)}{k^2 {(k-p)}^2 {(l+p)}^2 {(l+k)}^2{(l+k+p)}^2}
\Big],
\eea
where $S(p)=S(p_1,p_2)|_{p_1=0}$ and $p= p_1$. 
Let us analyze the limit $p \to 0$. Since the numerator has a $p^2$
and the denominator is proportional to at most $1/p^2$, 
Eq.(\ref{sp}) has zeroth leading order in $p$, and $S(p)|_{p\to
  0}\equiv S$ is constant. Another reason for this is the following
one. If we omit all noncommutative factors the result is not singular at 
$p=0$ since it is of zeroth order in $1/p$. If we introduce 
noncommutativity, additional infrared singularities can arise if and
only if the supergraph is divergent \cite{Arefeva}. However, this supergraph
is evidently ultraviolet finite, hence there is no infrared
singularity in it (notice that the external momentum $p$ plays the
role of an infrared cutoff). 

The constant $S$ can be written as 
$S=S_{pl}+S_{np}$ where $S_{pl}$ is a planar contribution to $S$ given
by the first two terms in Eq.(\ref{sp}), and $S_{np}$ is a nonplanar
contribution given by the two last terms. It is evident that all
$p$-dependent cosine factors cannot decrease the power of
$p$. Therefore if we can set $p=0$ in all cosines  (but not in
denominator!) it will not change the infrared behavior. 
Let us find the leading contributions at $p\to 0$ to effective action from
the second term of Eq.(\ref{sp}).  
After using the Feynman representation and integration over $l$ we
arrive at 
\bea
& &
\label{pl}
\frac{\lambda^5}{8}
\frac{p^2}{(4\pi)^2}\int\frac{d^4 k}{(2\pi)^4}\int_0^1 dx dy
\frac{1}{k^2(k+p)^2}e^{-2ik\times p x}\times\\&\times&
\frac{\sqrt{(p^2 x +k^2 y +(k+p)^2 (1-x-y)-(px+ky+(k+p)(1-x-y))^2)
    p\circ p}}{ p^2 x +k^2 y +(k+p)^2 (1-x-y)- (px+ky+(k+p)(1-x-y))^2}
\times\nonumber\\&\times&
K_{-1}\Big(\sqrt{(p^2 x +k^2 y +(k+p)^2
  (1-x-y)-(px+ky+(k+p)(1-x-y))^2)p\circ p}\Big).\nonumber
\eea
Here $K_{-1}(z)$ is the modified Bessel function of order $-1$. 
Let us consider this expression in limit $p\to 0$. Remind that
$K_{-1}(x)\sim \frac{1}{4x}+O(x)$ for $x\to 0$
(we do not use the explicit form of $O(x)$ since it corresponds to terms
proportional to $p^4$ which are not essential for our purposes). 
We find that this expression has the same $p\to 0$
limit as  
\bea
\label{pl1}
& &\frac{\lambda^5}{32}
\frac{p^2}{(4\pi)^2}\int\frac{d^4 k}{(2\pi)^4}\int_0^1 dx dy
\frac{1}{k^2(k+p)^2}\times\\
&& \frac{1}{p^2 x +k^2 y
    +(k+p)^2(1-x-y)-(px+ky+(k+p)(1-x-y))^2}+O(p^4). \nonumber
\eea
The term containing the noncommutative factor vanishes. 
Hence this contribution in leading order
is equal to $\frac{6}{(4\pi)^4}\zeta(3)$ which could have been
obtained if we had set $\cos(p\times l)=1$ from the very beginning.
As a result, the sum of first two terms of Eq.(\ref{sp}), which
corresponds to the planar correction in the limit $p\to 0$, has the
contribution 
\bea
\label{s1}
\frac{\lambda^5}{24} p^2
\int\frac{d^4 k d^4 l}{{(2\pi)}^8}
\frac{1}{k^2 {(k+p)}^2 l^2
  {(l+k)}^2{(l+k+p)}^2}=\frac{\lambda^5}{2(4\pi)^4}\zeta(3). 
\eea
We used the expression for this integral given in \cite{Buch3}. As
pointed out before there is no noncommutative contribution to this
result. Hence the total contribution to the chiral effective action
from the planar sector is 
\bea
\label{pl2}
{\cal L}^{(c)}_{pl}=\frac{\lambda^5}{4(4\pi)^4}\zeta(3)\int d^6 z
\Phi^3+O(\Phi^2\Box^2\Phi). 
\eea

It remains to find out the nonplanar contribution to the chiral
effective potential given by the two last terms of Eq.(\ref{sp}).
We can use the Feynman representation and then integrate over $k$ with
the help of the identity given in \cite{Alv} to find 
\begin{equation}
\label{np}
S_{np}=\frac{\lambda^5}{48}\frac{p^2}{32\pi^2}\int\frac{d^4
  l}{(2\pi)^4}\frac{1}{l^2(l+p)^2} G(l|p),
\end{equation}
where
\bea \label{(40)}
G(l|p)&=& 
\int_0^1 dx dy e^{-2i p\times l y} 
K_{-1} \Big(\sqrt{[p^2 x +(p-l)^2 y -(px+(p-l)y)^2]l\circ l}\Big)
\times\nonumber\\&\times&
\frac{\sqrt{4[p^2 x +(p-l)^2 y -(px+(p-l)y)^2]l\circ l}}{p^2 x +(p-l)^2
  y -(px+(p-l)y)^2}+\nonumber\\&+& 
\int_0^1 dx dy e^{-2ip\times l (x+y)}
K_{-1} \Big(\sqrt{[l^2 x +(p+l)^2 y -(lx+(p+l)y)^2]l\circ l}\Big)
\times\nonumber\\&\times&
\frac{\sqrt{4[l^2 x +(p+l)^2 y -(lx+(p+l)y)^2]l\circ l}}{l^2 x +(p+l)^2
  y -(lx+(p+l)y)^2}.
\eea
This is the exact two loops result for the nonplanar contribution to
the chiral effective potential. 

The integral in the right hand side of Eq.(\ref{(40)}) is very
complicated. To estimate such an integral we use the following
approximation. Let us rewrite the integral in the form
\begin{equation}
\label{np1}
S_{np}= \frac{\lambda^5}{48}
\Big[\frac{p^2}{32\pi^2}\int_{0}^{\Lambda^2}\frac{d^4 l}{(2\pi)^4} 
\frac{G(l|p)_{small}}{l^2(l+p)^2}+
\frac{p^2}{32\pi^2}\int_{\Lambda^2}^{\infty}\frac{d^4 l}{(2\pi)^4} 
\frac{G(l|p)_{large}}{l^2(l+p)^2}\Big],
\end{equation}
where $\Lambda$ is an arbitrary scale. We use the notation 
$G(l|p)_{small}$ and $G(l|p)_{large}$ to mean that for the corresponding
interval we take the asymptotic form of the function $G(l|p)$ at small
and large arguments, respectively.  
Since the modified Bessel function $K_{-1} (x)$ has the asymptotic behavior
$K_{-1}(x)\sim\frac{1}{4x}+O(x)$ 
for small $x$ and
$K_{-1}(x)\sim(-\sqrt{\frac{\pi}{2x}}+O(\frac{1}{x}))e^{-x}$
 for large $x$, we have for $p$  small
\bea
G(l|p)_{small}&=&
\int_0^1 dx dy e^{-2i p\times l y} 
\frac{1}
{p^2 x +(p-l)^2 y-(px+(p-l)y)^2} 
+\nonumber\\&+&
\int_0^1 dx dy e^{-2ip\times l (x+y)}
\frac{1}{l^2 x
  +(p-l)^2 y -(lx+(p-l)y)^2}+\ldots, 
\eea
and
\bea
G(l|p)_{large}&=&
\int_0^1 dx dy  
\exp(-a l^2)
\frac{\sqrt[4]{[p^2 x +(p-l)^2 y -(px+(p-l)y)^2]l\circ l}}{p^2 x
  +(p-l)^2 y -(px+(p-l)y)^2}+\nonumber\\&+& 
\int_0^1 dx dy 
\exp(-a l^2)
\frac{\sqrt[4]{[l^2 x +(p+l)^2 y -(lx+(p+l)y)^2]l\circ l}}{p^2 x
  +(p-l)^2 y -(px+(p-l)y)^2}.
\eea
Due to the asymptotics of $K_{-1}(x)$ at small values of the argument
  we can see that the next-to-leading term in its expansion (it is of
  first order in the argument) can lead only to contributions
  proportional to $p^4$.

We use the same choice for $\theta_{\mu\nu}$ as before
(see the discussion which lead to Eq.(\ref{det})). For small
$p$ we then get
\bea
S_{np}&=&\frac{\lambda^5}{48}\Big[
\frac{p^2}{32\pi^2}\int_0^{\Lambda^2}\frac{d^4 l}{(2\pi)^4}
\frac{1}{l^2(l+p)^2(p^2x+(p-l)^2y+(px+(p-l)y))^2}
+\nonumber\\&+&
\frac{p^2}{32\pi^2}\int_{\Lambda^2}^{\infty}\frac{d^4 l}{(2\pi)^4}
\exp(-a l^2)\frac{1}{l^6}\Big].
\eea 
The first integral  
can be approximated as 
\bea
\frac{\lambda^5}{8(4\pi^2)^2}\zeta(3)-
\frac{\lambda^5}{96}\frac{p^2}{(4\pi)^4}\Lambda^2, 
\eea
where we took into account that $\int_0^{\Lambda^2}=\int_0^{\infty}
-\int_{\Lambda^2}^{\infty}$ and approximated 
$(l^2(l+p)^2(p^2x+(p-l)^2y+(px+(p-l)y))^2)^{-1}$, in the last interval,
as $l^{-6}$ for small $p$. The second term  
can be calculated straightforwardly. It is equal to
$\frac{p^2a}{2(16\pi^2)^2}\beta(a\Lambda^2)$
where 
$\beta(a\Lambda^2)=\int_{a\Lambda^2}^{\infty}\frac{dz}{z^{3/2}}e^{-z}$.
Then the total nonplanar contribution is 
\bea
\label{np2}
S_{np}=\frac{\lambda^5}{4(16\pi^2)^2}\Big[\zeta(3)+
p^2a\beta(a\Lambda^2)-\frac{p^2}{24\Lambda^2}\Big].
\eea
For $p<<l$ the argument of $K_{-1}$ in Eq.(\ref{np}) is  $l^2
a$. Since the border between the two asymptotic forms of the Bessel
function is $l^2 a =1$, then, it is natural to choose $\Lambda$
satisfying $\Lambda^2 a =1$, that is, $\Lambda = \frac{1}{\sqrt{a}}$. 
Hence we get
\bea
\label{np3}
S_{np}=\frac{\lambda^5}{4(4\pi^2)^2}[\zeta(3)+(\beta-\frac{1}{24}) p^2a ].
\eea 
where $\beta=\beta(a\Lambda^2)|_{\Lambda=\frac{1}{\sqrt{a}}}\simeq
0.178$.
The corresponding contribution to the effective action is 
\bea
{\cal L}^{(2)}_{np}=\frac{\lambda^5}{4(4\pi^2)^2}\int d^6 z
[\zeta(3)\Phi^3+(\beta-\frac{1}{24}) a\Phi^2\Box\Phi ].
\eea
The noncommutative effects arise in the terms proportional to
$\int d^6 z\Phi^2(a\Box)\Phi $. A natural interpretation is the
following. Let us suppose that the external momentum
$p$ is very small but non-zero. This suggests that the
noncommutativity parameter $a$ may be very large.
Then we find that at $a p^2 \sim 1$
(or as is the same $a\Box\Phi\sim\Phi$) 
we have sizable corrections to effective action which do not
vanish at small energy. 
Therefore, the total contribution from the planar and nonplanar parts
to the low-energy effective action is 
\bea
\label{sfin}
{\cal L}^{(2)}=\frac{\lambda^5}{2(16\pi^2)^2}\zeta(3)\int d^6 z \Phi^3
+\frac{\lambda^5}{4(16\pi^2)^2}(\beta-\frac{1}{24}) a\int d^6 z
\Phi^2\Box\Phi+O(\Phi^2\Box^2\Phi).
\eea
This correction  is finite and does not require any 
renormalization.  It is evident that it reproduces
the known results for Wess-Zumino model
\cite{Buch4,Buch3,West2,West3} at the commutative limit $a\to 0$.

It is interesting to point out that the second term in Eq.(\ref{sfin}) may
equivalently be rewritten as an integral over full superspace with the
help of identity 
\begin{equation}
\int d^6 z\Phi^2 \Box \Phi = \frac{1}{16}\int d^6 z\Phi^2 \bar{D}^2
D^2 \Phi =-\frac{1}{4}\int d^8 z \Phi^2 D^2 \Phi.
\end{equation} 
As a result, the quantum correction under consideration may be represented
by a local functional either in full 
superspace or in chiral superspace. None of these representations
is preferable. However, since such a correction depends only on $\Phi$
and can be written as a local functional over chiral superspace 
it is natural to refer to it as a contribution to the chiral effective
action.

To conclude, we have calculated the leading chiral 
correction to the superfield effective action in the noncommutative
Wess-Zumino model. It is finite and does not possess any singularity
coming from the UV/IR mixing. We found that this
correction contains a standard part which coincides with the two loops
chiral effective potential in the commutative Wess-Zumino model and
terms depending on $p^2 a$, where $p$ plays the role of an energy
scale and $a$ is the noncommutativity parameter.  In the standard
case we set $p\to 0$, however, if we have very strong
noncommutativity, that is $a\to\infty$, we obtain non-trivial
corrections in Eq.(\ref{sfin}) at $p^2a\to const$. The presence of such
a correction can be related to the 
quantum dynamics of the vacuum in which fluctuations of geometry are
correlated with the energy of the particles created.

We have also calculated the one and two loops contributions to the
K\"{a}hlerian effective potential. This is the first calculation of
higher loop contributions 
to the effective action in a noncommutative supersymmetric field theory
carried out with the use of supergraph techniques. This approach allows us
to preserve manifest supersymmetry at    
all steps of the calculation. In the one loop K\"ahlerian
effective potential all dependence on the noncommutativity parameter
vanishes, and the result coincides with the commutative case
\cite{Buch3}. It is natural to expect the same result for the one loop
K\"ahlerian effective potential in any noncommutative theory. The
two loops K\"ahlerian effective potential has a planar part which has the same
form as in the commutative case \cite{Buch4}, and a nonplanar part 
which is strongly dependent on the noncommutativity. 
It turns out that if the noncommutativity is large, the nonplanar
contribution is suppressed by fast oscillations of the 
nonplanar term. Otherwise, if the noncommutativity is small, 
the nonplanar contribution becomes the leading one. 

{\bf Acknowledgements.} The authors are grateful to S. M. Kuzenko for
useful comments.
The work of I.L.B. is partially supported by INTAS, project
INTAS-991-590, project INTAS-001-254 and RFBR, project No. 99-02-16617. 
The work of M.G. was partially supported by CNPq and FAPESP. 
The work of A.Yu.P. is supported by FAPESP, project No. 00/12671-7. 
The work of V.O.R. is partially
supported by CNPq and PRONEX under contract CNPq 66.2002/1998-99.

\end{document}